# Nonvolatile Memory Cells Based on MoS$_2$/Graphene Heterostructures

Simone Bertolazzi, Daria Krasnozhon, Andras Kis[*]

Electrical Engineering Institute, École Polytechnique Fédérale de Lausanne (EPFL), CH-1015 Lausanne, Switzerland

*Correspondence should be addressed to: Andras Kis, andras.kis@epfl.ch

**ABSTRACT**

Memory cells are an important building block of digital electronics. We combine here the unique electronic properties of semiconducting monolayer MoS$_2$ with the high conductivity of graphene to build a 2D heterostructure capable of information storage. MoS$_2$ acts as a channel in an intimate contact with graphene electrodes in a field-effect transistor geometry. Our prototypical all-2D transistor is further integrated with a multilayer graphene charge trapping layer into a device that can be operated as a nonvolatile memory cell. Because of its band gap and 2D nature, monolayer MoS$_2$ is highly sensitive to the presence of charges in the charge trapping layer, resulting in a factor of $10^4$ difference between memory program and erase states. The two-dimensional nature of both the contact and the channel can be harnessed for the fabrication of flexible nanoelectronic devices with large-scale integration.

Two-dimensional materials, such as graphene and single layers of boron nitride (BN) or molybdenum disulfide (MoS$_2$) are the thinnest known materials with electronic properties that can be advantageous for a wide range of applications in nanotechnology.[1-6] Being only one layer thick, they represent the ultimate limit of scaling in the vertical direction and could offer reduced power dissipation because of reduced short channel effects.[7] They can also be regarded as a complete material library containing all the components necessary for building electronic circuits in which insulating BN could act as the substrate and gate dielectric barrier,[6] graphene as an interconnect while MoS$_2$ or another 2D semiconductor could play the role of a semiconducting channel.[4] Because graphene is semimetallic, it could form the ideal contact to 2D semiconductors, capable of supporting large current densities,[8] exceeding $10^9$ A/cm$^2$ (ref 9). Furthermore, the lack of dangling bonds at the interface between graphene and 2D semiconductors would suppress the appearance of interface states and charge traps. As the graphene's work function can be electrostatically or chemically tuned,[10] it could also be adapted to a wide variety of 2D semiconductors with different work functions and band gaps. In this work, we demonstrate the integration of a 2D semiconductor in the form of monolayer MoS$_2$ with graphene electrodes. The basic device functions as a field-effect transistor, with ohmic contacts and performance comparable to similar devices with metal contacts.[4] Furthermore, this device can serve as a prototype for more advanced devices based on a 2D architecture. We demonstrate this by adding a charge trapping layer in the form of a multilayer graphene floating gate, resulting in a new heterostructure capable of operating as a nonvolatile memory cell.

The use of 2D materials could offer immediate practical advantages for the realization of memory devices based on the floating gate transistor structure.[11] In current flash memory technology, serious hurdles need to be overcome to advance device miniaturization, both in the lateral and vertical directions. Vertical scaling leads to reduction of the program/erase voltages but is limited by the requirement of charge retention on the floating gate, which sets the limit for a minimum tunneling oxide thickness (~5 nm). On the other hand, lateral scaling, driven by the quest for higher data storage capability, is seriously limited by the capacitive coupling between the drain electrode and the floating gate, which results in a longer penetration of the drain field in the transistor channel[12] as the device is scaled. Furthermore, interference between neighboring cells leads to an undesirable spread in the threshold voltages of the devices.[13] The effect of capacitive interference can be significantly diminished through the reduction of the floating gate thickness,[14] while the reduction of electrode thickness can diminish the coupling between the electrodes and the floating gate. Because of this, it is extremely interesting to investigate the suitability of 2D materials for use in memory devices where they could replace traditional choices for semiconducting channels, interconnects and charge trapping layers.

Figure 1 shows the structure of our memory device, composed of two transistors fabricated on the same monolayer $MoS_2$ flake placed over graphene strips acting as source and drain electrodes. A piece of multilayer graphene (MLG), 4-5 layers thick, separated by a 6 nm thick tunneling oxide layer ($HfO_2$) from the monolayer $MoS_2$ channel acts as the floating gate. We chose multilayer graphene as the floating gate because of its high work function (4.6 eV) which is not sensitive to the number of layers and results in a deeper potential well for charge trapping and improved charge retention. Furthermore, the low conductivity along the vertical c-axis suppresses detrimental ballistic currents across the floating gate[15] while a higher density of states ($> 4.4 \times 10^{13}$ $cm^{-2}eV^{-1}$) with respect to its single-layer counterpart ($8 \times 10^{12}$ $cm^{-2}eV^{-1}$)[16] would allow more charge to be stored on it resulting in a larger memory window. The floating gate is further capped by a 30 nm thick blocking oxide layer at the top. The conductivity of the $MoS_2$ channel depends on the amount of charge stored in the floating gate (FG) and is modulated by the voltage $V_{cg}$ applied to the control gate electrode (CG) which can also be used to vary the amount of charge stored on the floating gate.

The fabrication process includes three transfer steps.[17] Figure 1C shows the appearance of the device at several stages of the fabrication procedure. We start by growing graphene using chemical vapor deposition (CVD)[18] and transferring it onto a silicon substrate with a 270 nm thick thermally grown silicon-dioxide ($SiO_2$) layer. Graphene is patterned into stripes and contacted with metal leads. $MoS_2$ is then exfoliated on another $SiO_2$/Si substrate covered with a sacrificial polymer layer. Individual layers are detected using optical microscopy[19] and transferred[20] on top of the graphene stripes which form the source and drain electrodes in direct contact with monolayer $MoS_2$. Fabrication continues with the deposition of the tunneling oxide layer on top of the $MoS_2$/graphene heterostructure, in the form of a 6 nm thick $HfO_2$ film grown using atomic layer deposition (ALD). We use a thermally oxidized Al seed layer[21] to facilitate the ALD deposition over graphene and $MoS_2$. Multilayer graphene (MLG) flakes 1.5 nm thick are transferred onto the tunneling oxide and positioned above the $MoS_2$ flakes.

Finally, the floating gate is capped by a 30nm thick HfO$_2$ layer followed by the deposition of the control gate.

Figure 2a shows the small-bias current vs bias voltage ($I_{ds}$ - $V_{ds}$) characteristic of one of our floating gate transistors acquired for different values of the control gate voltage $V_{cg}$. The symmetry of the curves with respect to the origin and their linear behavior indicates the formation of ohmic contacts between graphene and monolayer MoS$_2$. This is due to a tunable and favorable graphene work function.[10] The relatively large current, on the order of 500 nA for a 40 mV bias, corresponding to a total series resistance of 125 kOhm shows that efficient charge injection from graphene to MoS$_2$ is possible thanks to the large overlap of graphene 2p$_z$ with Mo4$d_{z^2}$ states responsible for electrical conduction in MoS$_2$. The use of graphene as a contact material for 2D semiconductors in place of thicker metallic films is expected to be advantageous as it allows fabricating devices and circuits[22] in a truly 2D architecture, using for example roll-to-roll printing and with reduced parasitic interference between neighboring devices and increased resistance to short-channel effects.[7] Work-function tunability[10] of graphene will allow it to be adapted to a wide variety of 2D materials. Furthermore, because it is chemically inert and mechanically flexible, graphene is the ideal non-invasive contact to other 2D materials.

The information storage capability of our device can be deduced from the transfer characteristics (drain current $I_{ds}$ vs. control-gate voltage $V_{cg}$), shown in figure 2b, acquired in two sweeping directions, from negative to positive voltages (blue) and in the opposite sweep direction (red). The large hysteresis, characterized by a maximum voltage shift $\Delta V$ of approximately 8 V, is related to the charging/discharging of the floating gate in response to the control gate voltage. Application of a positive voltage to the control gate ($V_{cg} > 0$) induces Fowler-Nordheim electron tunneling from the channel into the floating gate through the HfO$_2$ oxide barrier, figure 2c.[23] Consequently, the threshold voltage of the device is shifted towards higher positive voltages and device current is decreased. This corresponds to the program state of the device characterized by a low drain current $I_{ds}$. On the other hand, when the control gate voltage is negative, electrons are pushed back from the floating gate to the channel and for sufficiently high electric fields the floating gate is fully discharged, resulting in the erase state, characterized by high $I_{ds}$. For the program and erase states we measured a maximal program/erase (P/E) current ratio exceeding $10^4$. Such a remarkable P/E current ratio not only allows easy readout of the device state, but also allows multilevel storage, where more than one bit of information could be stored in the memory cell in the form of several distinct floating-gate charge levels.

Other hysteresis mechanisms[24] could be present in the device, but they can be ruled out as potential sources of such a high memory effect. Trapping states related to charge impurities present at the semiconductor/dielectric interface cannot generate a permanent threshold voltage shift. As shown in figure 3a, we can generate permanent current/threshold voltage states characterized by varying memory windows $\Delta V$ by modulating the amount of charge stored in the floating gate (FG). In order to calculate the threshold voltage shift, we define a reference erase state by assuming that the floating gate is not likely to accumulate positive charge (see Supporting Information). The floating gate can be depleted of all residual charge by sweeping the control gate voltage from 0 V to a low negative voltage, denoted as $V_{cg,min}$ (< -5 V). This places the device in the erase state. Immediately after floating gate depletion, we sweep the control gate voltage in the negative-to-positive direction, starting from $V_{\text{cg, min}}$ = -18 V, resulting in the black dashed line in figure 3a, characterized by a threshold voltage $V_{\text{th}} \approx 0\text{V}$. Repeating this voltage sweep results in a set of curves with no observable dependence of the $V_{\text{th}}$ on the different values tested for $V_{\text{cg,min}}$.[17] This indicates that the multilayer graphene floating gate is not likely to accumulate positive charge (holes) in response to a negative voltage applied to the control gate. Hence, a reference $E$ state with $V_{\text{th}} \approx 0$ V could be defined, corresponding to approximately zero charge stored in the FG. We can now reach different program (P) states by performing successive $V_{\text{cg}}$ sweeps, stopping at gradually increasing maximum values of $V_{\text{cg,max}}$. Figure 3a shows a family of curves acquired in this way, corresponding to different program states characterized by their own memory windows. The threshold voltage shift $\Delta V$ (memory window), measured from the reference $E$ state, is strongly dependent on the maximum control gate voltage, as can be inferred from the inset graph in figure 3a. A maximum control gate voltage of +18 V results in a memory window exceeding 8 V.

We estimate the amount of charge stored on the floating gate from the expression[25] $n = \left( \Delta V \times C_{FG\text{-}CG} \right) / q$ where $q$ is the electron charge, $\Delta V$ is the threshold voltage shift and $C_{FG\text{-}CG}$ is the capacitance between the floating gate and the control gate, modeled as $C_{FG\text{-}CG} = \varepsilon_0 \varepsilon_{bl} / d_{bl}$ with $\varepsilon_0$ the vacuum permittivity, $\varepsilon_{bl}$ the relative dielectric constant of the $HfO_2$ blocking layer (~25) and $d_{bl}$ its thickness (~30 nm). This results in a density of stored electrons on the order of ~$3.7 \times 10^{13}$ cm$^{-2}$. In agreement with Mishra et al.[14] and Hong et al.[26] we conclude that MLG, due to its higher density of states than in graphene, allows a larger memory window with strong potential for multilevel data storage.

We now test the dynamic behavior of our device, reported in figure 3b. Switching between program and erase states is achieved through the application of voltage pulses to the control gate electrode, with the source grounded and the drain biased at 50 mV. The device is initially in the $E$ state, corresponding to the ON current level. A positive voltage pulse (+18 V) with the duration of 100 ms leads to both accumulation of charge in the transistor channel, indicated by the ~150 nA current spike and to tunneling of electrons into the floating gate. When the control gate voltage is reset to 0 V, the device remains in its program state, holding a stable OFF current. The application of a symmetric negative pulse (-18 V, 100 ms duration) 2.3 seconds later restores the initial erase state. We performed an endurance test measurement for our memory device, observing a well-defined and reproducible $P/E$ switching for over 120 cycles. The slow decay of the ON current is presumably due to trapping mechanisms related to charge impurities present at the semiconductor/dielectric interface, as observed elsewhere.[24]

The stability of $E$ and $P$ states, crucial for non-volatile information storage, was further investigated by monitoring their time-resolved behavior at a constant drain-source bias of 50 mV, reported in Figure 4. During the entire observation time, the corresponding erase state current was measured to be in the range $10^{-8}$ -$10^{-7}$ A, with an exponential decay saturating to ca. 20 nA. After this measurement, the memory was programmed to the program state through a voltage pulse ($V_{cg}$ = +18 V, duration 3 s), resulting in the OFF state of the drain current $I_{ds}$. Although slowly increasing with time, the $I_{ds}$ current remained in the $10^{-12}$ -$10^{-10}$ A range during at least 2000 seconds. Room for improvement can be identified in both lowering the density of charge impurities in and around the channel in order to stabilize the erase-state current, and in improved engineering of the tunneling oxide layer to increase the retention performance. As the final step of our device characterization procedure, we studied the memory charge retention characteristics (figure 4b), which are related to both the nature of the tunneling barrier and to the work function of the MLG. The latter determines the depth of the potential well for the electrons stored in the floating gate. After programming the memory through a positive pulse ($V_{cg}$ = +18 V during 3 sec), we measured the threshold voltage at different time intervals, determined using a linear fit of the transfer characteristics, as shown in the inset of figure 4b. These curves were acquired in a narrow range of control voltage $V_{cg}$, from 0 V to +12 V, in order to prevent undesired electron depletion/injection occurring at the floating gate. Successive voltage sweeps carried out immediately one after the other resulted in similar $V_{th}$ values, confirming the fact that the measurement itself was not perturbing the system. We observe a threshold voltage variation from 7.5 V to 5 V during a time period of $10^4$ seconds. Assuming that the reference erase state is characterized by $V_{th} \approx 0$ V, we estimate that after 10 years, 30% of the initial charge would still be stored on the floating gate. This result proves that the charge leakage from the floating gate is slow on the time-scale of years and that the combination $MoS_2$/MLG indeed has a potential for application in non-volatile memory technology. The value extracted for charge retention appears encouraging, considering that this is the first prototype of its kind and further improvement can be achieved by means of a thorough engineering of the blocking oxide layer, using for instance novel insulating 2D crystals, such as BN or 2D oxides.[27]

In conclusion, we have demonstrated a new type of heterostructures based on 2D materials in which we use graphene as an ohmic contact to monolayer $MoS_2$ in a field-effect transistor geometry, paving the way towards the realization of truly 2D device architectures. The final device includes a multilayer graphene floating gate and operates as a non-volatile memory cell. This demonstrates that it is possible to design memory devices using 2D building blocks, including contacts, floating gate and the semiconducting channel. Moreover, the excellent mechanical properties of 2D semiconductors such as $MoS_2$ (ref 28) can be exploited for fabrication of transistor circuits and memory devices on flexible substrates, with a naturally emerging range of related applications. Such devices could be produces massively and inexpensively using liquid-scale processing[29] or roll-to-roll printing[30] of CVD-grown material.[31,32]

## MATERIALS AND METHODS

Single layers of $MoS_2$ are exfoliated from commercially available crystals of molybdenite (SPI Supplies Brand Moly Disulfide) using the scotch-tape micromechanical cleavage technique method pioneered for the production of graphene.[1] AFM imaging was performed using the Asylum Research Cypher AFM. Electrical characterization of the memory device was performed in ambient conditions at room temperature using an Agilent E5270B parameter analyser.

## ACKNOWLEDGMENTS

Device fabrication was carried out in part in the EPFL Center for Micro/Nanotechnology (CMI). Thanks go to Zdenek Benes for help with e-beam lithography. We thank Branimir Radisavljevic and Jacopo Brivio for valuable discussions and Mahmut Tosun for technical help with CVD graphene growth. This work was financially supported by the Swiss National Science Foundation (grants no. 138237 and 135046) and the Swiss Nanoscience Institute (NCCR Nanoscience).

**REFERENCES**

1   Novoselov, K. S. *et al.* Electric Field Effect in Atomically Thin Carbon Films. *Science* **306**, 666-669, (2004).
2   Novoselov, K. S. *et al.* Two-dimensional atomic crystals. *PNAS* **102**, 10451-10453, (2005).
3   Dean, C. R. *et al.* Boron nitride substrates for high-quality graphene electronics. *Nature Nanotech.* **5**, 722-726, (2010).
4   Radisavljevic, B., Radenovic, A., Brivio, J., Giacometti, V. & Kis, A. Single-layer $MoS_2$ transistors. *Nature Nanotech.* **6**, 147-150, (2011).
5   Britnell, L. *et al.* Field-Effect Tunneling Transistor Based on Vertical Graphene Heterostructures. *Science*, (2012).
6   Britnell, L. *et al.* Electron Tunneling through Ultrathin Boron Nitride Crystalline Barriers. *Nano Lett.* **12**, 1707-1710, (2012).
7   Yoon, Y., Ganapathi, K. & Salahuddin, S. How Good Can Monolayer MoS2 Transistors Be? *Nano Lett.* **11**, 3768–3773, (2011).
8   Moser, J., Barreiro, A. & Bachtold, A. Current-induced cleaning of graphene. *Appl. Phys. Lett.* **91**, 163513, (2007).
9   Yu, J., Liu, G., Sumant, A. V., Goyal, V. & Balandin, A. A. Graphene-on-Diamond Devices with Increased Current-Carrying Capacity: Carbon sp2-on-sp3 Technology. *Nano Lett.* **12**, 1603-1608, (2012).
10  Yu, Y.-J. *et al.* Tuning the Graphene Work Function by Electric Field Effect. *Nano Lett.*, (2009).
11  Kahng, K. & Sze, S. M. A floating gate and its application to memory devices. *Electron Devices, IEEE Transactions on* **14**, 629-629, (1967).
12  Chan, A. C. K. *et al.* SOI flash memory scaling limit and design consideration based on 2-D analytical modeling. *Electron Devices, IEEE Transactions on* **51**, 2054-2060, (2004).
13  Jae-Duk, L., Sung-Hoi, H. & Jung-Dal, C. Effects of floating-gate interference on NAND flash memory cell operation. *Electron Device Letters, IEEE* **23**, 264-266, (2002).
14  Misra, A. *et al.* in *Memory Workshop (IMW), 2012 4th IEEE International.* 1-4.
15  Raghunathan, S., Krishnamohan, T., Parat, K. & Saraswat, K. in *Electron Devices Meeting (IEDM), 2009 IEEE International.* 1-4.
16  Zhu, W., Perebeinos, V., Freitag, M. & Avouris, P. Carrier scattering, mobilities, and electrostatic potential in monolayer, bilayer, and trilayer graphene. *Physical Review B* **80**, 235402, (2009).
17  info, S.
18  Li, X. *et al.* Large-Area Synthesis of High-Quality and Uniform Graphene Films on Copper Foils. *Science* **324**, 1312-1314, (2009).
19  Benameur, M. M. *et al.* Visibility of dichalcogenide nanolayers. *Nanotechnology* **22**, (2011).
20  Brivio, J., Alexander, D. T. L. & Kis, A. Ripples and Layers in Ultrathin $MoS_2$ Membranes. *Nano Lett.* **11**, 5148-5153, (2011).
21  Kim, S. *et al.* Realization of a high mobility dual-gated graphene field-effect transistor with $Al_2O_3$ dielectric. *Applied Physics Letters* **94**, 062107-062103, (2009).

22  Radisavljevic, B., Whitwick, M. B. & Kis, A. Integrated circuits and logic operations based on single-layer $MoS_2$. *ACS Nano* **5**, 9934–9938, (2011).

23  Lenzlinger, M. & Snow, E. H. Fowler-Nordheim Tunneling into Thermally Grown $SiO_2$. *Journal of Applied Physics* **40**, 278-283, (1969).

24  Late, D. J., Liu, B., Matte, H. S. S. R., Dravid, V. P. & Rao, C. N. R. Hysteresis in Single-Layer $MoS_2$ Field Effect Transistors. *ACS Nano* **6**, 5635-5641, (2012).

25  Pavan, P., Larcher, L. & Marmiroli, A. *Floating gate devices : operation and compact modeling.* (Kluwer Academic, 2004).

26  Hong, A. J. *et al.* Graphene Flash Memory. *ACS Nano* **5**, 7812-7817, (2011).

27  Osada, M. & Sasaki, T. Two-Dimensional Dielectric Nanosheets: Novel Nanoelectronics From Nanocrystal Building Blocks. *Adv. Mater.* **24**, 210-228, (2012).

28  Bertolazzi, S., Brivio, J. & Kis, A. Stretching and Breaking of Ultrathin $MoS_2$. *ACS Nano* **5**, 9703-9709, (2011).

29  Coleman, J. N. *et al.* Two-Dimensional Nanosheets Produced by Liquid Exfoliation of Layered Materials. *Science* **331**, 568-571, (2011).

30  Bae, S. *et al.* Roll-to-roll production of 30-inch graphene films for transparent electrodes. *Nature Nanotech.* **5**, 574-578, (2010).

31  Liu, K.-K. *et al.* Growth of Large-Area and Highly Crystalline $MoS_2$ Thin Layers on Insulating Substrates. *Nano Lett.* **12**, 1538-1544, (2012).

32  Zhan, Y., Liu, Z., Najmaei, S., Ajayan, P. M. & Lou, J. Large-Area Vapor-Phase Growth and Characterization of $MoS_2$ Atomic Layers on a $SiO_2$ Substrate. *Small* **8**, 966-971, (2012).

**FIGURES**

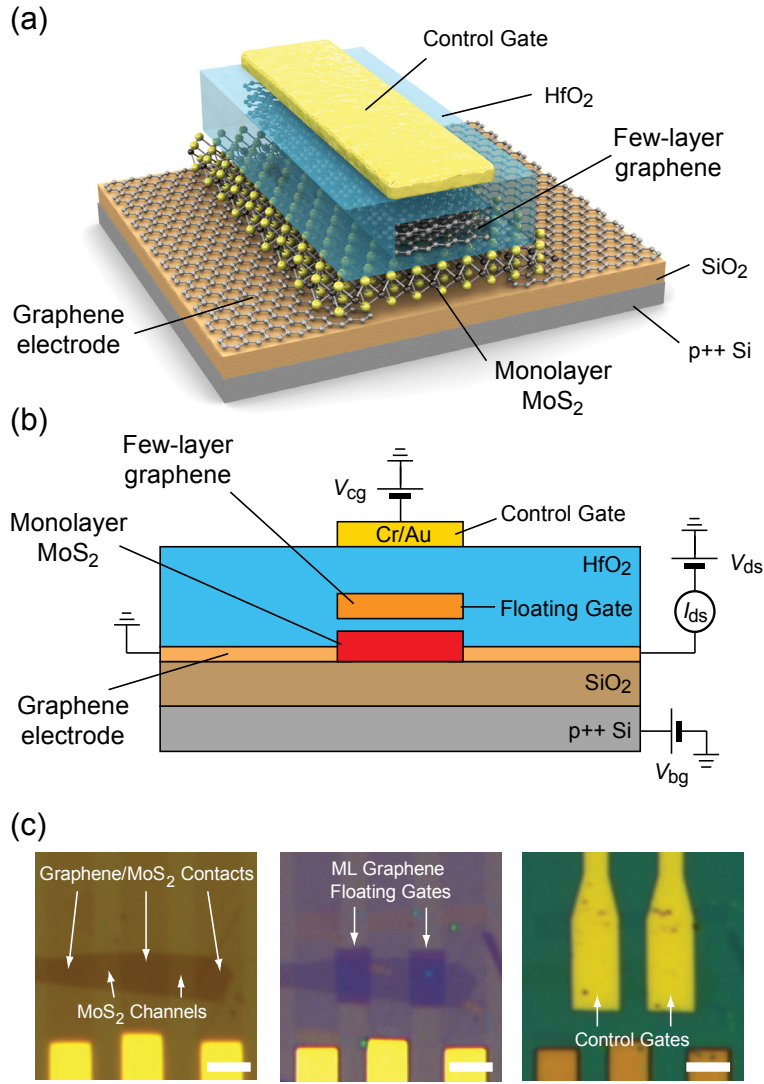

**Figure 1. MoS$_2$/graphene heterostructure memory layout.** (a) Three-dimensional schematic view of the memory device based on single-layer MoS$_2$ (b) Schematics of a heterostructure memory cell with a single-layer MoS$_2$ semiconducting channel, graphene (SLG) contacts and multilayer graphene (MLG) floating gate. The MLG floating gate is separated from the channel by a thin tunneling oxide (cca. 1 nm Al$_2$O$_3$ + 6 nm HfO$_2$) and from the control gate by a thicker blocking oxide (1 nm Al$_2$O$_3$ + 30 nm HfO$_2$). (c) Optical micrographs of two graphene/MoS$_2$ heterostructure transistors fabricated on the same MoS$_2$ monolayer flake at various stages of fabrication. Left: single-layer MoS$_2$ transferred over an array of graphene stripes patterned on oxidized silicon chips and contacted with metal leads (Cr/Au: 10/50 nm); Middle: the same device after deposition of the tunneling oxide and transfer/patterning of MLG floating gates; Right: final device after the deposition of the blocking oxide and definition of the control gate electrodes (Cr/Au: 10/50 nm). Scale bars: 3 μm.

(a)

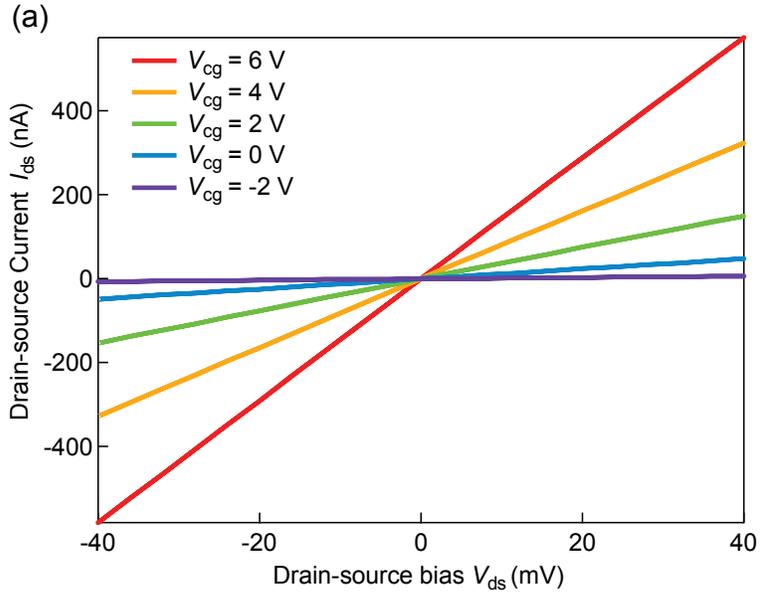

(b)

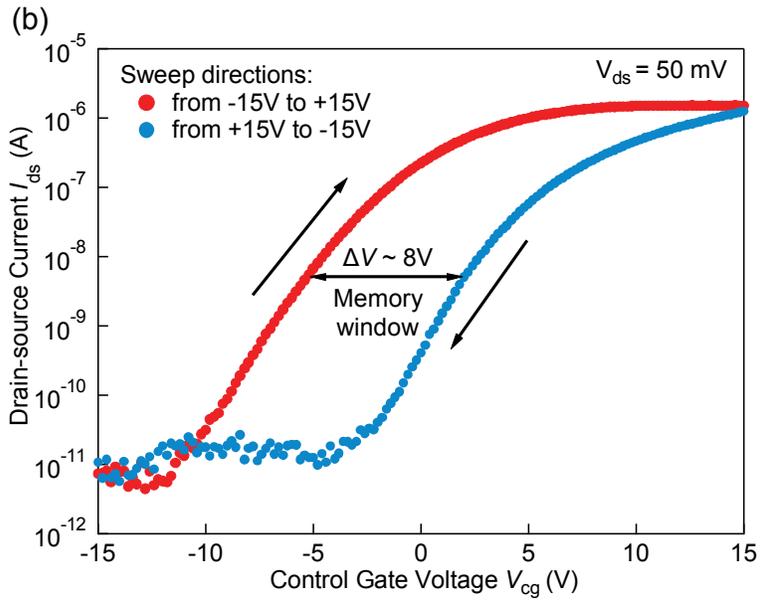

(c)

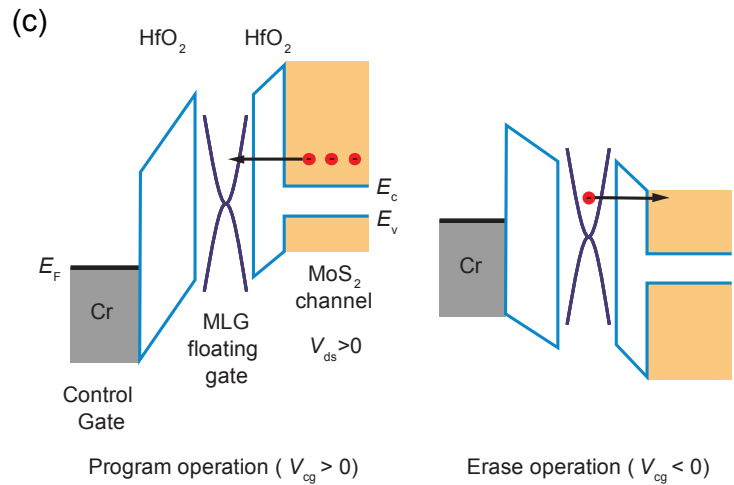

**Figure 2. Device characteristics and charge trapping in the device.** (a) Output characteristics (drain-source current $I_{ds}$ vs. drain-source voltage $V_{ds}$) of the floating gate transistor in the ON state, for different control gate biases $V_{cg}$. The curve linearity at small $V_{ds}$ indicates that graphene acts as a ohmic contact to monolayer $MoS_2$. (b) Transfer characteristic (drain-source current $I_{ds}$ vs. control-gate voltage $V_{cg}$) of the floating gate transistor, acquired along two different control-gate voltage sweep directions. The large hysteresis of ~8 V is related to accumulation of charge in the MLG floating gate. (c) Simplified band diagram of the memory device in program and erase states. A positive control gate voltage $V_{cg}$, corresponding to the program state of the device. Electrons tunnel from the $MoS_2$ channel through the 6 nm thick $HfO_2$ and accumulate on the multilayer graphene floating gate. Application of a negative control gate voltage $V_{cg}$ depletes the floating gate and resets the device.

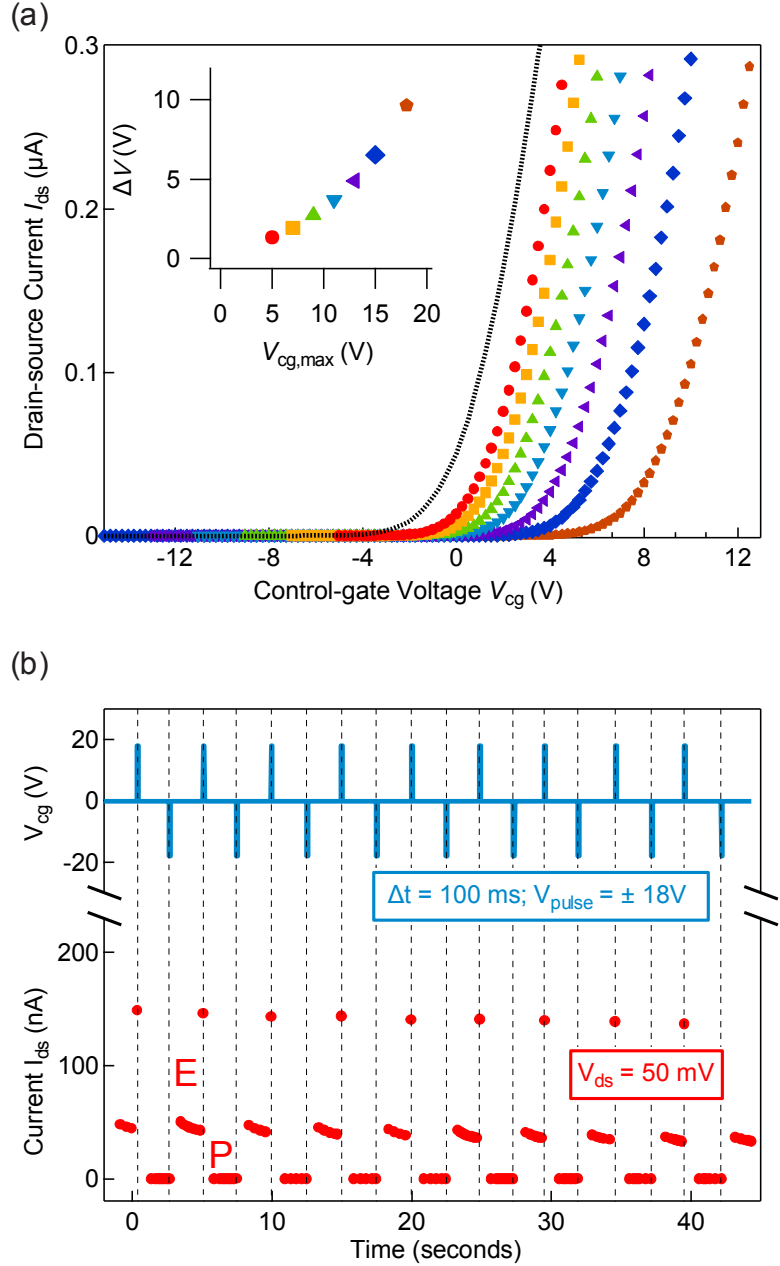

**Figure 3. Memory states.** (a) Dependence of the memory window on the control-gate voltage. Symbol lines: transfer characteristics (drain-source current $I_{ds}$ vs. control-gate voltage $V_{cg}$) of the floating gate transistor acquired from positive to negative control gate voltages $V_{cg}$, for different $V_{cg,max}$ (plotted in the inset). The drain-source voltage $V_{ds}$ is fixed to 50 mV. Dashed line: transfer curve of the device in the erase state, corresponding to no charge stored on the floating gate. The inset graph shows the dependence of the memory window – related to charging of the floating gate – on the maximum voltage applied to the control gate. (b) Switching between erase (*E*, high current, device ON) and program (*P*, low current, device OFF) states induced by the application of alternating $V_{cg}$ pulses (± 18 V for 100 ms) with a time interval of 2.3 s. The application of a positive $V_{cg}$ pulse (E→P) induces a drain-source current $I_{ds}$ peak (130-150 nA) due to the increase of the charge density in the MoS$_2$ channel during the pulse.

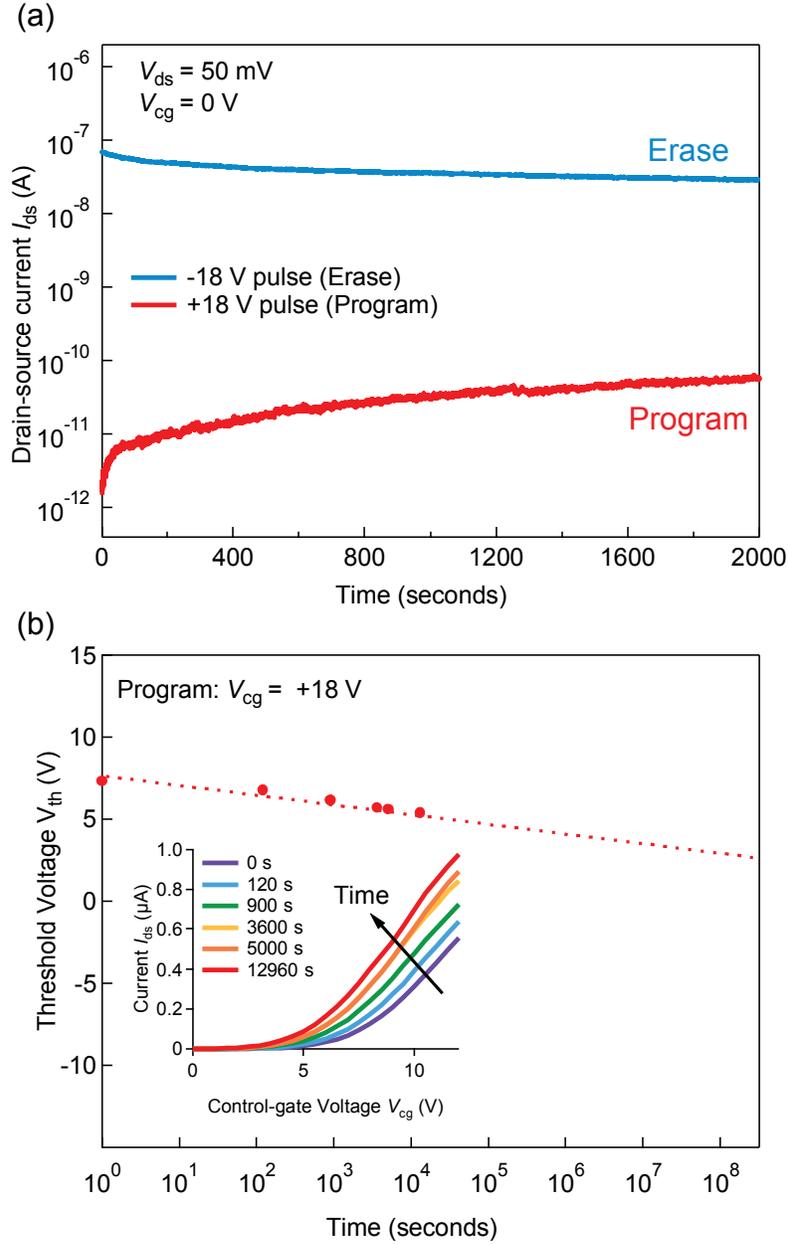

**Figure 4. Device dynamics and charge retention.** (a) Temporal evolution of drain-source currents ($I_{ds}$) in the erase (ON) and program (OFF) states. The curves are acquired independently for the program ($10^{-8}$ – $10^{-7}$ A) and erase ($10^{-12}$ – $10^{-10}$ A) current states and plotted on a common time scale. The drain-source bias voltage is 50 mV and the duration of the control-gate voltage ($V_{cg}$) pulse is 3 seconds. (b) Time-resolved behavior of the device threshold voltage after application of the programming voltage pulse ($\pm$ 18 V for 3 seconds). Threshold voltage was determined from linear fits to transfer curves (inset) in the linear regime. We estimate that after ten years the device retains 30% of the charge stored on the floating gate.

# Supplementary information for Nonvolatile Memory Cells Based on MoS₂/Graphene Heterostructures

Simone Bertolazzi, Daria Krasnozhon, Andras Kis[*]

Electrical Engineering Institute, École Polytechnique Fédérale de Lausanne (EPFL), CH-1015 Lausanne, Switzerland

*Correspondence should be addressed to: Andras Kis, andras.kis@epfl.ch

## DEVICE FABRICATION

The fabrication process includes three transfer steps, involving CVD graphene, mechanically exfoliated single-layer MoS₂ and multilayer graphene (MLG). We started by employing the chemical vapor deposition technique to grow graphene on 99.8% pure copper foils (Alfa Aesar) according to the two-step growth recipe reported in Li et al.[1] CVD graphene is transferred[2] onto oxidized silicon chips (thermally grown dry oxide, SiO₂ thickness ≈ 270 nm; silicon resistivity < 0.005 Ω·cm) and patterned with e-beam lithography and O₂ plasma etching in the shape of an array of stripes, 3 μm wide, interspaced by a 2 μm distance (Figure S1 a). These stripes are used as the injecting/collecting contacts for the transistors and are contacted with metal leads (Cr/Au) fabricated with e-beam lithography, e-beam evaporation and lift-off procedure. A longitudinally shaped monolayer MoS₂ flake (ca. 10 to 15 μm long, 2.5 μm wide) is moved and aligned over the graphene strip array, with its longitudinal axis perpendicular to the stripes, as shown in figure S1b. Care has been taken in avoiding and/or removing short-circuiting paths among the graphene stripes due to additional MoS₂ flakes participating in the transfer.

The transfer procedure for single-layer $MoS_2$ was conducted as follows. First, thin flakes of $MoS_2$ were exfoliated from naturally occurring crystals of molybdenite (SPI supplies) onto $SiO_2$/Si substrates covered with a double polymer layer of polyvinyl alcohol (PVA) and polymethyl methacrylate (PMMA). The sample surface is imaged using an optical microscope (Olympus BX51 M) to detect candidate monolayer $MoS_2$ flakes. In previous work based on atomic force microscopy (AFM) we established an optical-contrast/thickness relationship for $MoS_2$ flakes lying on $SiO_2$ (ref [3]) and PVA/PMMA coated silicon substrates.[4] These results were used to identify $MoS_2$ monolayers flakes on our polymer-coated samples and subsequently for verification of the thickness after transferring them onto $SiO_2$/Si substrates. Before transfer, the flake thickness and the $MoS_2$ surface topography were further investigated by AFM.

Once an interesting monolayer $MoS_2$ flake is selected, the PMMA film is peeled off the $SiO_2$/Si chip using dry transfer[5] and laid down, with the $MoS_2$ flakes facing up, on a transparent polydimethyl siloxane (PDMS) support, which is used for alignment and release of the flake onto the target substrate. With the help of a micromanipulator we aligned the PDMS/PMMA-supported monolayer $MoS_2$ over the array of graphene stripes (see above). During the alignment, the sample temperature was maintained at 90 °C. When the polymer film touches the sample surface, the temperature is increased to 120 °C; since the PMMA film sticks preferentially to the $SiO_2$ surface, the PDMS support can be easily lifted off, releasing the PMMA/flakes on the substrate. The samples are then immersed in acetone to strip the PMMA and annealed in a vacuum furnace at 400°C, for 4 hour in Ar/H$_2$ atmosphere. The result of the MoS$_2$ transfer procedure is shown in figure S1b.

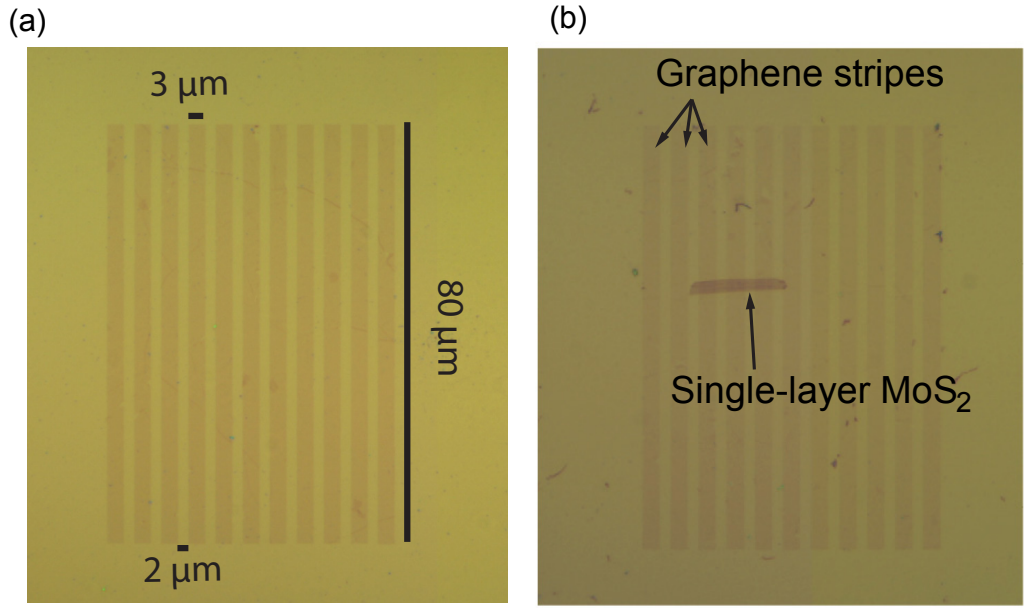

**Figure S1.** (a) Array of CVD graphene stripes patterned on a SiO$_2$/Si chip. (b) Single-layer MoS$_2$ flake transferred onto the array with its longitudinal axis perpendicular to the stripes. Two transistors in series could be fabricated, with channels defined by the gap between the stripes.

The fabrication continues with the deposition of the tunneling oxide layer, which consists of a 6 nm thick ALD grown HfO$_2$ film. It is known that ALD growth of oxide films on graphene is not straightforward, due to the absence of dangling bonds on the surface of a defect-free graphene layer, which are required for the chemical reactions of the ALD process to occur. In order to obtain uniform coverage of both the graphene contacts and the MoS$_2$ channels, we thermally evaporated a seed layer of Al (<1 nm)[6] and let it oxidize in air, at 120 °C, for several hours. ALD deposition is then performed in a Beneq system using a reaction of H$_2$O with tetrakis(ethyl-methylamido)hafnium at 200°C.

MLG flakes, obtained by the micromechanical cleavage technique, are transferred as described in Brivio et al.[7] onto the tunneling oxide, over the MoS$_2$ flakes. The thickness of the selected MLG flakes was measured by AFM to be approximately 1.5 nm, i.e. 4-5 layers thick.

## DETERMINATION OF THE REFERENCE ERASE STATE

Depleting the floating gate with different values of the control gate voltage did not result in systematic change of the threshold voltage of the device. This suggests that negative control-gate voltages are effective in depleting the floating gate from stored electrons, but are not responsible for positive charge to be stored in the floating gate.

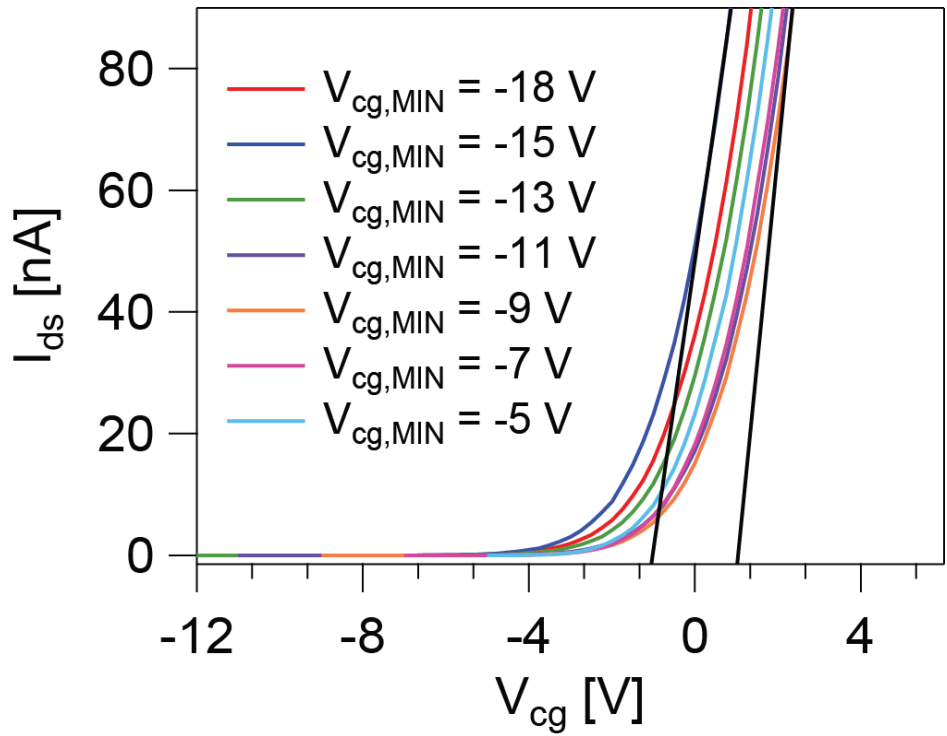

**Figure S2.** Transfer characteristics of the device in the E state, acquired after depleting the floating gate at $V_{cg} = V_{cg,MIN}$, for different values of $V_{cg,MIN}$. The threshold voltage is randomly distributed in a range from -1 V to +1V.

**BAND DIAGRAM OF THE FLOATING GATE TRANSISTOR**

For the construction of the band diagram of our floating gate transistor in the direction perpendicular to the gate stack, we assumed that the tunneling and the blocking oxide layers are made exclusively by ALD $HfO_2$. From previously reported electronic and morphological structure of thin $HfO_2$ films grown on $SiO_2$/Si, [8-10] we extract an approximate value for the energy band-gap of the order of 5.5 ± 0.4 eV (depending on film capping conditions) and for the electron affinity of ca. 2.5 eV. N-type single-layer $MoS_2$ has a Fermi level shifted towards the bottom of the conduction band. The exact position of the conduction band with respect to the vacuum level was obtained by adding the energy gap (1.8 eV) to the value of the photoelectric threshold (5.97 eV) calculated with the GGA + PAW method.[11] Hence, we consider the electron affinity of single-layer $MoS_2$ to be approximately 4.2 eV, which is close to the measured value for multilayer films. The work function for MLG was assumed to be equal to 4.6 eV (ref [12]), considering a MLG floating gate with a number of layers > 4L. The barrier in the conduction band between the $HfO_2$ tunneling oxide and $MoS_2$ is then equal to $\phi_b = \chi_{HfO_2} - \chi_{MoS_2} = 1.7 \ eV$ , which allows for lower programming voltages, compared to $SiO_2$ tunneling oxides. The depth of the potential well at the floating gate is calculated from the formula: $\phi_w = \chi_{HfO_2} - \phi_{MLG} = 2.1 \ eV$.

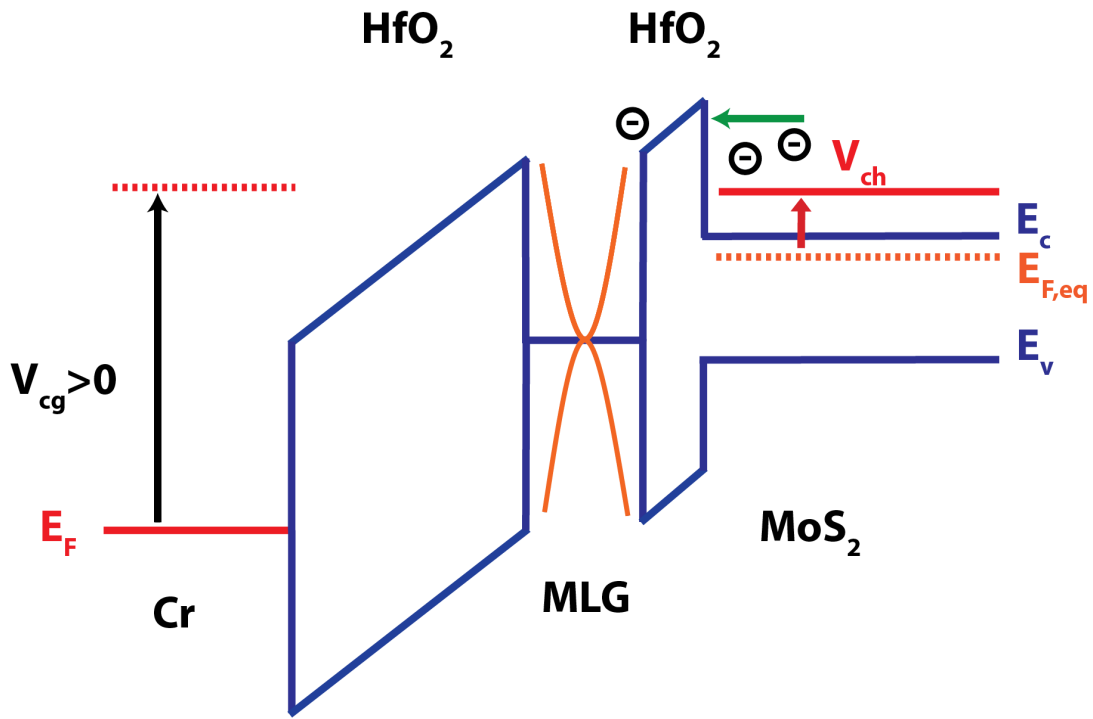

**Figure S3.** Band diagram of the floating gate transistor in the direction perpendicular to the gate stack, when a positive voltage is applied to the control-gate electrode and no charge is stored in the floating gate. See text for energy barrier values. Under a positive $V_{cg}$ bias condition, the Fermi level of MoS$_2$ shifts into the conduction band, resulting in a significant accumulation of electrons ($\sim 10^{13}$ cm$^{-2}$) in the transistor channel. The total voltage drop between the control-gate electrode (biased at $V_{cg}$) and the semiconducting channel (connected to the ground) is given by the algebraic sum of the potential drops in the HfO$_2$ tunneling and blocking oxide layers, plus the potential drop $V_{ch}$ due to the quantum capacitance of the ultrathin semiconducting layer (for simplicity we suppose no band offsets). For high control-gate voltages $V_{cg}$ the tunneling oxide energy barrier becomes steeper and narrower, allowing Fowler-Nordheim to occur between MoS$_2$ and MLG.

## GRAPHENE CONTACTS

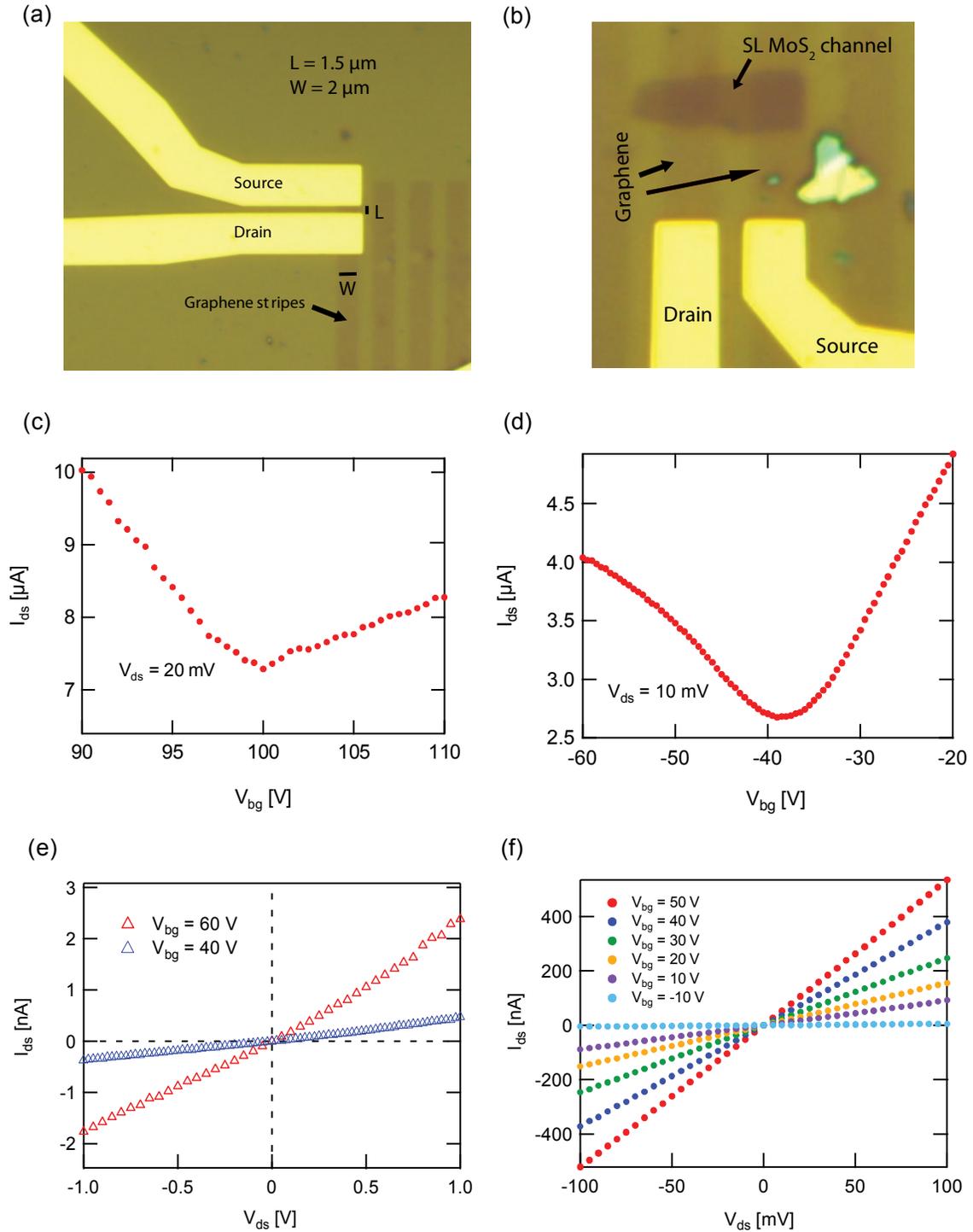

**Figure S4.** (a) Optical micrograph of the test transistor built on the prepatterned CVD graphene stripes on SiO$_2$/Si, used to extract information on the electronic properties (e.g. doping) of the graphene electrodes. (b) Optical micrograph of a back-gated transistor, with graphene injecting/collecting contacts. The channel width is ca. 3 µm and the channel length, given by the interspacing between the stripes is 2 µm. (c) Transfer characteristic (back-gate sweep) of the graphene transistor shown in (a), acquired after source and drain contact lift-off. (d) Transfer characteristic of the same device, after deposition of $HfO_2$ using an Al seed layer. The neutrality point is now at negative gate voltages, indicating n-type doping. (e,f) Output characteristic of the back gated graphene transistors, before (e) and after (f) the deposition of the $HfO_2$ film using an Al seed layer.

As-fabricated graphene test transistors presented a high positive neutrality point (> +40 V), indicating strong p-type doping, which was presumably induced during different fabrication and transfer steps. Although electron injection effectively occurred from the p-doped graphene into the single-layer $MoS_2$ flake, the corresponding output characteristics are not symmetric, showing Schottky-barrier like behavior. Simultaneous gating of the graphene electrodes through the back-gate is not expected to be responsible for the observed asymmetric output characteristics: the work function of graphene, as measured by Kelvin Probe Microscopy, varies in the narrow range 4.5-4.8 eV as a response to the gating electric field.[13] We noticed that the deposition of a sub-nanometer thick film of Al and its oxidation into an $Al_2O_3$ seed layer for the growth of ALD $HfO_2$, results in a shift of the graphene neutrality point towards negative gate voltages (~-40 V).

## MEASUREMENT OF THE LEAKAGE CURRENT

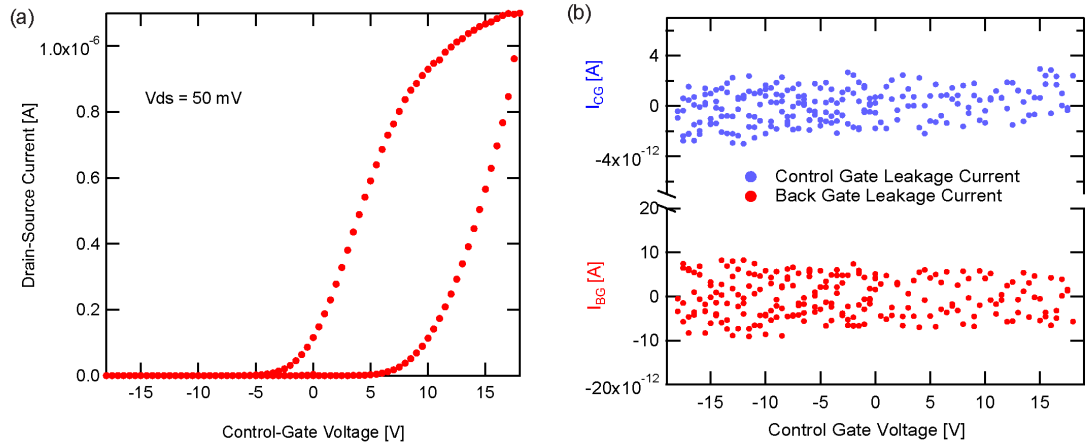

**Figure S5.** (a) Transfer characteristic ($I_{ds}$ vs. $V_{cg}$) of the floating gate memory and (b) simultaneous measurement of the gate leakage currents at the back gate (grounded) and control gate electrodes.

## REFERENCES

1   Li, X. *et al.* Graphene Films with Large Domain Size by a Two-Step Chemical Vapor Deposition Process. *Nano Lett.* **10**, 4328-4334, (2010).

2   Li, X. *et al.* Large-Area Synthesis of High-Quality and Uniform Graphene Films on Copper Foils. *Science* **324**, 1312-1314, (2009).

3   Benameur, M. M. *et al.* Visibility of dichalcogenide nanolayers. *Nanotechnology* **22**, (2011).

4   Bertolazzi, S., Brivio, J. & Kis, A. Stretching and Breaking of Ultrathin $MoS_2$. *ACS Nano* **5**, 9703-9709, (2011).

5   Dean, C. R. *et al.* Multicomponent fractional quantum Hall effect in graphene. *Nat Phys* **7**, 693-696, (2011).

6   Kim, S. *et al.* Realization of a high mobility dual-gated graphene field-effect transistor with $Al_2O_3$ dielectric. *Applied Physics Letters* **94**, 062107-062103, (2009).

7   Brivio, J., Alexander, D. T. L. & Kis, A. Ripples and Layers in Ultrathin MoS2 Membranes. *Nano Letters* **11**, 5148-5153, (2011).

8   Cheynet, M. C., Pokrant, S., Tichelaar, F. D. & Rouviere, J.-L. Crystal structure and band gap determination of $HfO_2$ thin films. *Journal of Applied Physics* **101**, 054101, (2007).

9   Sayan, S., Garfunkel, E. & Suzer, S. Soft x-ray photoemission studies of the $HfO_2/SiO_2/Si$ system. *Applied Physics Letters* **80**, 2135-2137, (2002).

10  Kukli, K., Ritala, M., Lu, J., Haårsta, A. & Leskelä, M. Properties of HfO2 Thin Films Grown by ALD from Hafnium tetrakis(ethylmethylamide) and Water. *Journal of The Electrochemical Society* **151**, F189-F193, (2004).

11  Ataca, C. & Ciraci, S. Functionalization of Single-Layer MoS2 Honeycomb Structures. *J. Phys. Chem. C* **115**, 13303-13311, (2011).

12  Hibino, H. *et al.* Dependence of electronic properties of epitaxial few-layer graphene on the number of layers investigated by photoelectron emission microscopy. *Physical Review B* **79**, 125437, (2009).

13  Yu, Y.-J. *et al.* Tuning the Graphene Work Function by Electric Field Effect. *Nano Letters* **9**, 3430-3434, (2009).